\documentclass{jfm}
\usepackage{graphicx}
\usepackage{amsmath,amssymb}
\usepackage{epstopdf, epsfig}
\usepackage[normalem]{ulem}

\usepackage{float}
\usepackage{bm}

\usepackage{color}
\usepackage{mathtools}
\def\strutdepth{\dp\strutbox}
\def\nw#1{\strut\vadjust{\kern-\strutdepth\vtop to0pt{\vss\hbox to\hsize
{\hskip\hsize\hskip5pt$\leftarrow$\hss\strut}}}{\em #1}}

\shorttitle{Film deposition of a self-propelled droplet on a cone with slip}
\shortauthor{T. S. Chan, C. Pedersen, J. Koplik, A. Carlson}

\title{Film deposition and dynamics of a self-propelled wetting droplet on a cone with slip}

\author{Tak Shing Chan\aff{1} \corresp{\email{taksc@math.uio.no}}, Christian Pedersen \aff{1}, Joel Koplik \aff{2}, Andreas Carlson\aff{1} \corresp{\email{acarlson@math.uio.no}}}

\affiliation{\aff{1}Mechanics Division, Department of Mathematics, University of Oslo, Oslo 0316, Norway \aff{2} Benjamin Levich Institute and Department of Physics, City College of the City University of New York,
New York, New York 10031, USA
	}

\begin{document}
\maketitle

\begin{abstract}\label{gen:abstr}
We study the dynamic wetting of a self-propelled viscous droplet using the time-dependent lubrication equation on a conical-shaped substrate for different cone radii, cone angles and slip lengths. The droplet velocity is found to increase with the cone angle and the slip length, but decrease with the cone radius. We show that a film is formed at the receding part of the droplet, much like the classical Landau-Levich-Derjaguin (LLD) film. The film thickness $h_f$ is found to decrease with the slip length $\lambda$. By using the approach of matching asymptotic profiles in the film region and the quasi-static droplet, we obtain the same film thickness as the results from the lubrication approach for all slip lengths. We identify two scaling laws for the asymptotic regimes: $h_fh''_o \sim Ca^{2/3}$ for $\lambda\ll h_f$ and $h_f h''^{3}_o\sim (Ca/\lambda)^2$ for $\lambda\gg h_f$, here $1/h''_o$ is a characteristic length at the receding contact line and $Ca$ is the capillary number. We compare the position and the shape of the droplet predicted from our continuum theory with molecular dynamics simulations, which are in close agreement. Our results show that manipulating the droplet size, the cone angle and the slip length provides different schemes for guiding droplet motion and coating the substrate with a film.

\end{abstract}

\section{Introduction}\label{intro}
Coating a film onto a substrate as a liquid is forced to move along it is a
technique used in painting and industrial applications such as lithography, which has been studied since the early twentieth century \citep{Q99}. Dip coating is one way to coat a plate as it is withdrawn from a liquid reservoir above a critical plate velocity \citep{SDAF06,MALEKI2011359,gao2016}. A mathematical model describing this film coating was developed in the seminal work by \cite{LL42} and \cite{D43}. These theoretical works sparked a great interest in film coating later adopted for a range of solid geometries, \textit{e.g.} cylindrical fibres \citep{White1966,Wilson1988,deryck1996}, coating by rollers \citep{taylor1963,Wilson1982} and coating the inner surface of a channel/tube \citep{B61,Tabeling1986}. Some studies have focused on how other physical effects influence the film deposition such as gravity \citep{D43,SZAFE08}, inertia \citep{deryck1996,Orsini2017}, surfactants \citep{CARROLL197323}, particles on the interface \citep{dixit2013,homsy2013,Colosqui2013}, van der Waals forces for deposited films of nanometric scales \citep{Qu1989}, as well as effects of substrate roughness \citep{Homsy2005} and confinement due to the reservoir \citep{kim2017}. Much of the extensive literature on the film deposition dynamics have been summarized in several review articles (\cite{Ruschak1985,Q99,WeRu04,RIO2017100}).

The classical theory by \cite{LL42} and \cite{D43} gives a fundamental description of thin film coating, where the deposited film thickness is so thin that gravity can be neglected. The flow inside the film region is maintained by the balance of capillarity, characterized by the liquid/air surface tension coefficient $\gamma$, and the viscous forces, characterized by the liquid viscosity $\eta$. The film region is connected to a quasi-static liquid reservoir of a length scale that is much larger than the thickness of the deposited film. When a plate is withdrawn from a reservoir, this length is set by the capillary length $\ell_c\equiv (\gamma/\rho g)^{1/2}$, with $\rho$ the liquid density and $g$ the gravitational acceleration. By using the method of asymptotic matching, the thickness of the film $h_f$, denoted as the LLD film, is shown to have a universal scaling with respect to the plate velocity $U$ as $h_f/\ell_c\sim Ca^{2/3}$, where the Capillary number $Ca\equiv \eta U/\gamma$ is the ratio between the viscous and the surface tension forces. Remarkably, this $Ca^{2/3}$ power law has been demonstrated to be a robust relation in many different systems when a fluid film is deposited. The only required change in the scaling relation is to replace the capillary length by the corresponding characteristic length of the system. For example, the film thickness is rescaled by the fibre radius for the case when a cylindrical fibre is withdrawn from a bath \citep{White1966,DFJ74,Wilson1988}; while in the case of a long bubble moving in a tube, also known as Bretherton's problem, the corresponding length is the tube radius \citep{B61}.

Common to fluid coating processes is that they often require an external driving force to displace the fluid. However, when a droplet with a size smaller than the capillary length comes in contact with a conical fibre, it moves spontaneously from the tip to the base of the cone due to capillarity \citep{lorenceau2004,Er2013}. In nature, this self-propelled mechanism has been exploited by plants \citep{Lui2015} and animals \citep{Zheng2010,Wang9247} to facilitate water transport at small scales. When a droplet is translating above a critical velocity, a layer of liquid film is expected to be deposited on the conical surface at the receding part of the droplet. It has been discovered recently on the trichome of the {\em Sarrancenia} that the deposited film provides a wetted surface, enabling later water droplets to be transported at a velocity several orders of magnitude larger than found in other plants \citep{Chen2018}. Despite the importance of understanding the film deposition and potential implications for biological evolution in plants and giving a path to very fast droplet transport, the film deposition has not been studied before on conical geometries. Previous fluid coating studies have assumed a no-slip condition at the fluid-solid boundary, \textit{i.e.} no relative motion between the fluid and the solid boundary. Interestingly, slip lengths have been reported to be as large as a few micrometers for fluids such as polymer melts \citep{baeumchen09PRL} and for superhydrophobic surfaces \citep{Rothstein2010}. When the droplet size is decreased to a few micrometers or below, effects due to the fluid slip on solid surface may become significant as demonstrated in dynamical fluidic systems \citep{lauga07TXT,bocquetSCR09} such as the dewetting of microdroplets \citep{McGraw16,chan2017} and of liquid films \citep{FJMWW05}. However, the influence of slip on the droplet dynamics and film deposition is not known for the directional droplet motion on a cone.

 Although conical solid structures are common in nature and appear as a component in industrial processes, modeling of the droplet fluid flow on such geometries is lacking. In our recent article \citep{chan2020a}, we provide a physical picture of the spontaneous motion of the droplet based on the mismatch between the equilibrium contact angle and the apparent contact angles. This generates flow in the contact line regions and maintains the droplet motion. In this study, we implement the time-dependent lubrication equation developed in \cite{chan2020a} to investigate the evolution of the liquid-air interface of the capillary driven droplet motion on a smooth conical fibre. The properties of a deposited film generated by a self-propelled droplet are studied for small cone angles and for a wide range of slip lengths. Apart from the continuum approach, the simple geometry of a  conical shape allows us to study fluid flow using molecular dynamics (MD) simulations. Results from the MD simulations will be used as a verification for the droplet shape predicted by the lubrication model. In fact, the approach of MD simulations has previously been implemented to study wetting dynamics at the nanoscale \citep{nakamura13PRE}, the slip condition at a contact line region \citep{Qian2003PRE}, the frictional force on a sliding droplet \citep{Koplik2019} and the influence of physico-chemistry of water/substrate interface on droplet dynamics \citep{johansson2015}.

\section{Mathematical formulation}\label{form}
\begin{figure}
\begin{center}
\includegraphics[width=0.9\textwidth]{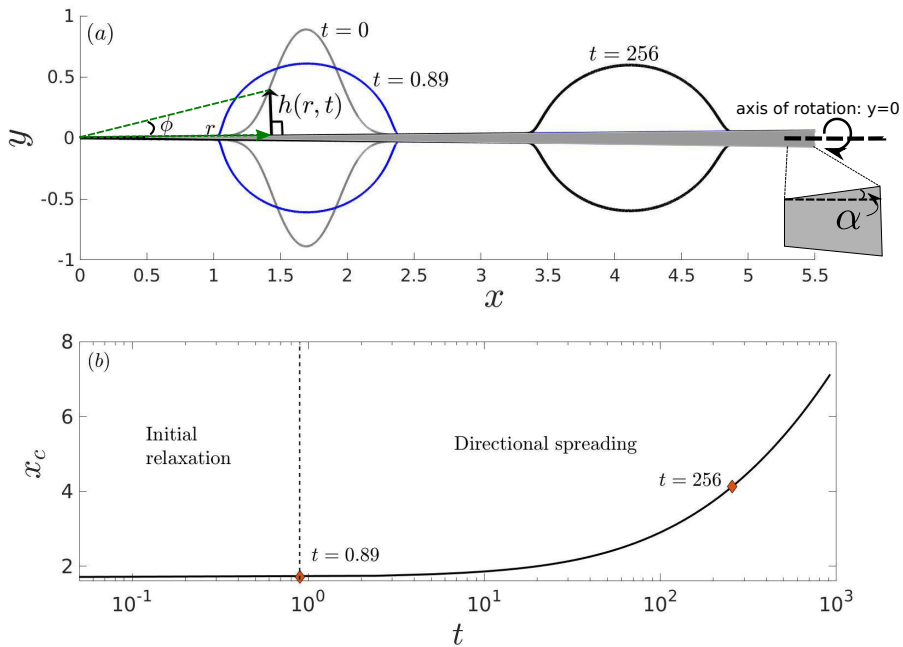}
\caption{(a) A description of the system at study, where a droplet is moving across (from left to right) a conically shaped fibre (grey shaded region). Droplet profiles $h(r,t)$ are shown at three different times on the fibre with a cone angle $\alpha$. From $t=0$ to $t=0.89$, there is a fast relaxation of the droplet before it slowly spreads across the fibre. (b) The center of mass of the droplet $x_c$ plotted as a function of time. The diamond markers correspond to the two profiles shown in (a), with a cone angle $\alpha=0.01$ rad, a prewetted film thickness $\epsilon=10^{-3}$ and a slip length $\lambda=0$.}\label{setup}
\end{center}
\end{figure}

An axisymmetric viscous droplet with a volume $V$ is placed in contact with a wetted surface of a conical fibre with a cone angle $\alpha\ll1$, see figure \ref{setup}(a). We consider a fibre surface prewetted with a thin layer of the same fluid of thickness $\epsilon$. The prewetted layer can be deposited or interpreted as a microscopic precursor film for a perfectly wetting droplet, \textit{i.e.} equilibrium contact angle $\theta_e=0^{\circ}$. The profile of the liquid-air interface is described by $h(r,t)$, the distance between the interface and the substrate, as a function of the distance from the vertex of the cone along its surface $r$ and time $t$. For droplets with a Bond number $Bo\equiv\rho g V^{2/3}/\gamma\ll1$, gravitational effects can be ignored. We consider the Reynolds number $Re\equiv\rho U V^{1/3}/\eta \ll1$ and the flow inside the droplet is described by the Stokes equations and the continuity equation.

\subsection{Lubrication approximation on a cone (LAC)}
Consider the flow in the droplet as $\vec{u}(r,\theta)$, here $\theta$ is
the polar angle measured from the axis of rotation. Supposing the polar angle
of the free surface of the droplet is very small, the flow is primarily in the radial direction. By using these approximations, the Stokes equations reduce to the lubrication equations here given in spherical coordinates \citep{chan2020a},
\begin{eqnarray}
	\label{g1}
	\frac{\partial p}{\partial r}=\frac{\eta}{r^2\theta}\frac{\partial}{\partial \theta}\left(\theta \frac{\partial u}{\partial \theta}\right),
\end{eqnarray}
\begin{eqnarray}
	\label{g2}
	\frac{\partial p}{\partial \theta}=0,
\end{eqnarray}
where $p$ is the pressure and $u$ is the radial velocity inside the droplet/film.

To describe the fluid flow (\ref{g1}-\ref{g2}) need to be accompanied by several boundary conditions.
At the liquid-air surface, the tangential stress is zero as we neglect viscous effects in the air
\begin{eqnarray}\label{bc1}
\frac{\partial u}{\partial \theta}=0\quad \mbox{at}\quad \theta=\alpha+\phi. \slabel{bc1}
\end{eqnarray}

At the wetted substrate, the normal velocity is zero and we assume a radial velocity described by the Navier-slip condition \citep{LBS05}
\begin{eqnarray}\label{bc2}
\frac{u}{\lambda}=\frac{1}{r}\frac{\partial u}{\partial \theta} \quad \mbox{at} \quad \theta=\alpha, \slabel{bc2}
\end{eqnarray}
where $\lambda$ is the slip length. 

%\begin{eqnarray}
%u=\frac{r^2\alpha^2}{2\eta}\frac{\partial p}{\partial r}\Bigg\{\left[\frac{\bar{\theta}^2-1}{2}-\left(1+\bar{\phi}\right)^2\ln\bar{\theta}\right]
%-\frac{\lambda}{r\alpha}\left[2\bar{\phi}+\bar{\phi}^2\right]\Bigg\},
%\label{f}
%\end{eqnarray}

Solving (\ref{g1}) and (\ref{g2}) with the boundary conditions (\ref{bc1}) and (\ref{bc2}) gives the velocity, and by imposing mass conservation of the liquid we get 
\begin{eqnarray}\label{c}
\frac{\partial h}{\partial t}+\frac{1}{r\alpha+ h}\frac{\partial}{\partial r} \Bigg[ && \frac{r^4\alpha^4}{2\eta}\frac{\partial p}{\partial r}\Bigg\{\frac{1}{8}\left[3\left(1+\frac{h}{r\alpha}\right)^4-4\left(1+\frac{h}{r\alpha}\right)^2+1\right] \nonumber \\
	&&-\frac{1}{2}\left(1+\frac{h}{r\alpha}\right)^4\ln(1+\frac{h}{r\alpha})-\frac{\lambda h^2}{2r^3\alpha^3}\left(2+\frac{h}{r\alpha}\right)^2\Bigg\} \Bigg]=0.
\end{eqnarray}
%\begin{eqnarray}
%&&G(h,r,\alpha,\lambda) \nonumber\\
%&&=\int_{1}^{1+\bar{\phi}}\alpha^2ur^2\bar{\theta} \mathrm{d}\bar{\theta} \nonumber\\
%&&=\frac{r^4\alpha^4}{2\eta}\frac{\partial p}{\partial r}\Bigg\{\frac{1}{8}\left[3\left(1+\bar{\phi}\right)^4-4\left(1+\bar{\phi}\right)^2+1\right] \nonumber \\
%&&-\frac{1}{2}\left(1+\bar{\phi}\right)^4\ln(1+\bar{\phi})-\frac{\lambda}{r\alpha}\left(2\bar{\phi}^2+\bar{\phi}^3\right)(1+\frac{\bar{\phi}}{2})\Bigg\} \nonumber \\
%\label{G}
%\end{eqnarray}

The pressure gradient inside the liquid is generated by the Laplace pressure $p=-\gamma\kappa$
where $\kappa$ is the curvature of the liquid-air interface, which for $\alpha \ll 1$ simplifies as
\begin{equation}\label{cur}
\kappa=\frac{h''}{(1+h^{'2})^{3/2}}-\frac{1-\alpha h'}{(r\alpha 
	+h )\left(1+h'^{2}\right)^{1/2}}
\end{equation}
with $()'\equiv \frac{\partial ()}{\partial r}$. 
The second term of the curvature is derived by using a rotation matrix with the cone angle $\alpha\ll 1$. We keep the $h'$ terms as the interface slope is not always small at the droplet scale. In the droplet region, we will see in section \ref{num_LAC} that the viscous effect is weak, and hence the droplet quickly adopts a quasi-static shape at the leading order, which is  determined by the uniform pressure condition i.e. $\kappa =$ constant. Although the flow field computed from the lubrication equation is inaccurate at the droplet scale, the correct quasi-static shape determines the flows in the contact line regions where lubrication approximation does work. Hence, Eq. (\ref{c}) is still valid for computing the evolution of the interface.

\subsection{Finite element method}\label{num_met}

We solve a coupled system of equations consisting of (\ref{c}) and the Laplace pressure equation $p=-\gamma\kappa$ numerically by using the finite element method. For the pressure equation, we split it into two following equations:
\begin{equation}\label{split1}
p=-\gamma\left[\frac{q'}{(1+h^{'2})^{3/2}}-\frac{1-\alpha h'}{(r\alpha 
	+h )\left(1+h'^{2}\right)^{1/2}}\right]
\end{equation}
and
\begin{equation}\label{split2}
q=h'.
\end{equation}
The variables we solve for are $h(r,t)$, $p(r,t)$ and $q(r,t)$. These fields are discretized with linear elements and solved as a coupled equation set by using Newton's method in the FEniCS library \citep{logg2012automated}. We use both an adaptive time stepping routine and an adaptive spatial discretization to refine the spatial resolution around the receding tail and the advancing front of the droplet with a resolution of  $\Delta r=10^{-4} V^{1/3}$, here $\Delta r$ is the difference of $r$ between two nodal points.
The numerical simulations are initialized with the initial profile $h(r,t=0)=\epsilon+A[1-\tanh(r-r_i)^2]$ where $A$ determines the volume of the droplet and $r_i$ determines the initial position of the droplet's geometric center. The simulations are insensitive to the initial droplet shape after a very short initial relaxation, see figure 1 and appendix  \ref{appenB}. Further, we impose the following boundary conditions at the boundary $\partial\Omega$ of the numerical domain: $p(r=\partial\Omega,t)=p(r=\partial\Omega,t=0)$ and $\partial h(r=\partial\Omega, t)/\partial r=$ 0.

\subsection{Molecular dynamics (MD) simulations}
We can test our hydrodynamic model by means of a ``numerical experiment'' - a
classical MD simulation of a liquid drop placed on a solid
cone, based on standard methods \citep{Frenfel02}. We
consider a generic viscous liquid consisting of spherically symmetric atoms
with a Lennard-Jones interaction, bound into linear tetramer molecules by a
FENE (finitely extensible nonlinear elastic) potential
\begin{equation}
V_{\rm LJ}(\mathcal{X}) = 4\,E\, \left[ \left( {\mathcal{X}\over\sigma} \right)^{-12} -
\left( {\mathcal{X}\over\sigma} \right)^{-6}\ \right]
\qquad
V_{FENE}(\mathcal{X})=-\frac{1}{2}\, k_F\, \mathcal{X}_0^2\, \ln\left(1-{\mathcal{X}^2\over \mathcal{X}_0^2}\right),
\end{equation}
where $\mathcal{X}$ is the separation between the center of mass of two atoms.
The LJ potential acts between all pairs of atoms within a cutoff distance
2.5$\sigma$, and is shifted by a linear term so that the force vanishes at
the cutoff. The FENE interaction (with parameters $k_F=30E/\sigma^2$
and $\mathcal{X}_0=1.5\sigma$, following \cite{Grest86} acts between adjacent
atoms on the chain.  The advantage of a molecular
rather than a monatomic liquid is that the vapor pressure is very low and
the liquid/vapor interface is relatively sharp and easy to visualize.
The solid is a conical section of a regular lattice whose atoms are mobile
but bound to their lattice sites by linear
springs with stiffness 100$E/\sigma^2$. The simulations are conducted
in an NVT ensemble, where the temperature is fixed at 0.8$E/k_B$
using a Nos\'e-Hoover thermostat.
This particular solid/liquid system has been used in a number
of previous simulations \citep{Busic03,Koplik2006,Koplik2013,Koplik2017}, and its properties
are well characterized. The liquid has bulk number density $0.857\sigma^{-3}$,
viscosity $5.18m/(\sigma\tau)$ and liquid-vapor surface tension
$0.668E/\sigma^2$, where $m$ is  the mass of 
the liquid atoms and $\tau=\sigma(m/E)^{1/2}$ is the
natural time scale based on the LJ parameters.  Furthermore, the liquid is
completely wetting:  a drop placed on flat solid surface with the same
density and interactions spreads completely into a thin film.

The simulation begins with all atoms on {\em fcc} lattice sites, within a
rectangular box of length 256.5$\sigma$ and sides 171$\sigma$, with repulsive
confining walls in the long ($x$) direction and periodic boundary conditions
on the sides. For the cone we
select all atoms in an fcc lattice of number density 1.06$\sigma^{-3}$ within a
radius $R_{co}(x)=3+x\tan\alpha$ (in the unit of $\sigma$) of the central axis,
which runs in the $x$-direction through the center of the box, and where
$\alpha=0.1$ radians. The resulting solid has 74,362 atoms.
The liquid initially occupies a disc-shaped region near the left edge,
$x_1<x<x_2$ and $R_{co}(x)<\mathcal{R}<R_{out}$ consisting of all atoms outside the cone but
inside an outer radius $R_{out}$. We have studied two cases (A,B) where the
remaining cone surface is initially dry and one case (C) where there is
also a liquid
(prewetting) film of thickness $4\sigma$. The parameters for the various
cases and the number of fluid atoms $N$ is given in Table I. The simulation
temperature starts at a low value in the solid phase and increases linearly
to the final value, $0.2\to 0.8E/k_B$ over 250$\tau$, to prevent the
liquid atoms from leaving the cone, and subsequently the drop is allowed
to evolve freely at the final temperature.

\begin{table}
\begin{center}
\begin{tabular}{|c||c|c|c|c|} \hline
case & $x_1$ & $x_2$ & $R_{out}$ & $N$    \\ \hline\hline
A    & 30    & 40    & 75  & 117880 \\ \hline
B    & 30    & 45    & 70  & 164120 \\ \hline
C    & 30    & 45    & 70  & 280280 \\ \hline
\end{tabular}
\caption{Initial geometry of the drops.}
\end{center}
\end{table}

\begin{figure}
\begin{center}
\includegraphics[width=1.\textwidth]{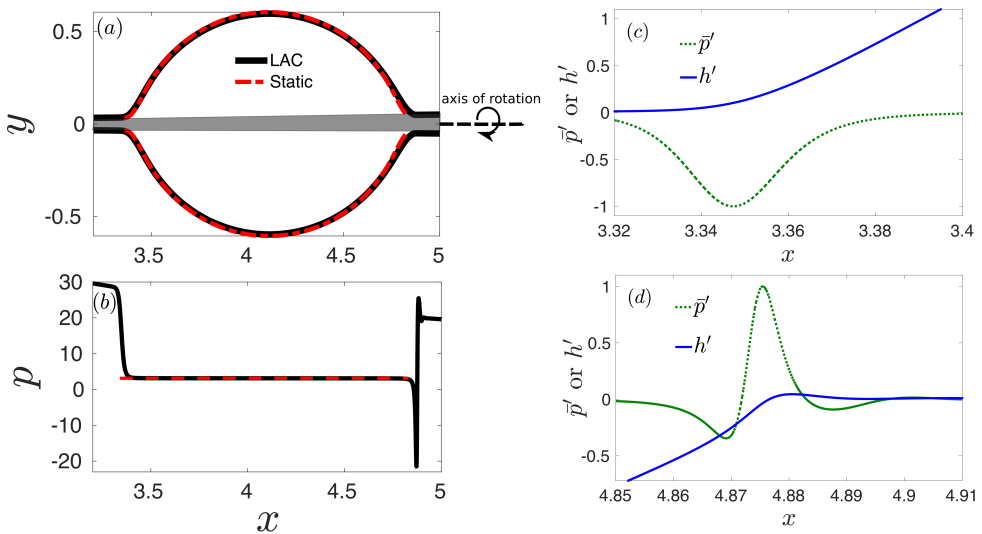}
\caption{(a) Solid line: Droplet shape on a conical fibre (grey shaded region) with a cone angle $\alpha=0.01$ rad at $t=256$ obtained from a numerical solution of the LAC with $\epsilon=10^{-3}$ and $\lambda=0$. Red dashed line: Static droplet shape obtained from solving the uniform curvature condition $\kappa=$ constant. (b) Solid line: The Laplace pressure $p$ as a function $x$ obtained from LAC. Red dashed line: The Laplace pressure of a static droplet, where the domain of the static droplet is between $x=3.36$ and $x=4.83$. (c) and (d) are the pressure gradient rescaled by its maximum value, denoted as $\bar{p}'$ and the interface slope $h'$  in the receding region in (c) and the advancing region in (d).}
\label{Fig2}
\end{center}
\end{figure}

\begin{figure}
\begin{center}
\includegraphics[width=1.2\textwidth]{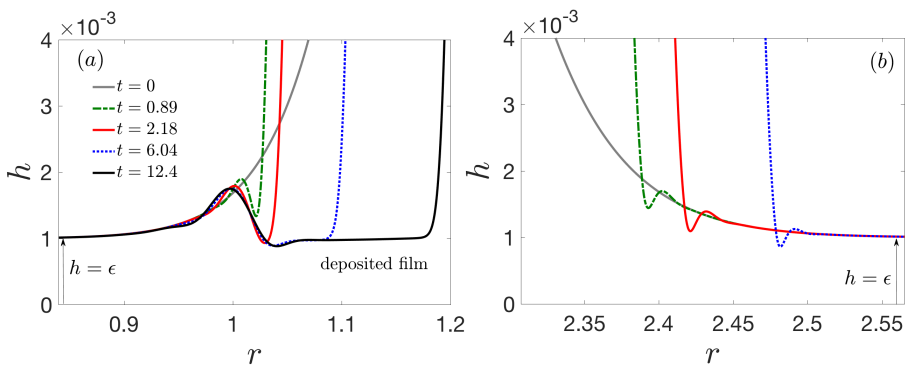}
\caption{(a) Interface dynamics at the receding region of the droplet, showing the formation of a deposited film from $t>2.18$. (b) The interface dynamics at the advancing region of the droplet. The far field conditions of the profiles at both ends match to a constant prewetted fluid layer of thickness $h=\epsilon = 10^{-3}$ and we have $\alpha=0.01$ rad and $\lambda=0$.}
\label{profiles_t}
\end{center}
\end{figure}

\section{Results and discussions}\label{res}
\subsection{Numerical solutions of the LAC}\label{num_LAC}
We first present numerical simulation results for the typical evolution of the droplet profile obtained from the lubrication approximation on a cone. In the following, all lengths are rescaled by $V^{1/3}$ and time is rescaled by $V^{1/3}\eta/\gamma$. The dimensionless parameters are the cone angle $\alpha$, the thickness of the prewetted layer $\epsilon$ and the slip length $\lambda$.

A typical dynamical process is shown in figure \ref{setup}(a). First the droplet relaxes from an initial shape ($t=0$) to a quasi-static shape ($t=0.89$) in a short time. At $t=0$, the initial shape gives a non-uniform curvature and hence a non-uniform pressure inside the whole droplet region. The pressure gradient generates flow inside the droplet. At $t=0.89$, a nearly uniform pressure distribution is achieved in the bulk of the droplet, but a large pressure gradient is created at the two edges of the droplet, which are commonly referred to as the $``$contact line regions", see figure \ref{Fig2}(b) for the pressure distribution at $t=256$. A concentration of stresses at the contact line would be expected (\cite{HS71}). After the quick initial relaxation, the droplet starts to propagate toward the thicker part of the cone. The position of the droplet is described by the center of mass of the droplet, for $\alpha\ll1$, defined as
\begin{equation}\label{xc}
x_c=\pi\int^{r_a}_{r_r}h(h+2\alpha r) rdr,
\end{equation}
which is plotted as a function of time in figure \ref{setup}(b). Here $r_r$ is the apparent receding and $r_a$ is the apparant advancing contact line positions, which are defined in the Appendix \ref{appenA}. The numerical simulations suggest that the droplet adopts a quasi-static shape during the directional spreading. For example, the profile for $t=256$ is plotted as the black solid line in figure \ref{Fig2}(a), and the pressure distribution is shown in figure \ref{Fig2}(b). Given the uniform pressure/curvature condition, one can solve for a static droplet profile, see the details of the computation in section \ref{asymp}. The static profile obtained in this way is plotted as the red dashed curve in figure \ref{Fig2}(a) for the same droplet position as obtained from the LAC at $t=256$. The agreement between the two profiles again demonstrates that the droplet profile is quasi-static on the droplet scale. As the droplet shape appears more round than flat, it may affect the validity of the lubrication approximation in the contact line regions where the viscous effects are significant. We zoom into the advancing and receding contact line regions and compute the pressure gradient $p'$ rescaled by its maximum magnitude and the interfacial slope $h'$, which are shown in figures \ref{Fig2}(c) and \ref{Fig2}(d). We observe that when approaching the contact line regions, the pressure gradient $p'$  increases from almost zero in the bulk of the droplet, and along with it the interface slope decreases. The maximum magnitude of the pressure gradient in both the receding and the advancing regions corresponds to an interfacial slope of magnitude less than 0.1 radians ($5.7^{\circ}$) which presumably fulfills the small slope assumption.

As shown in figures \ref{Fig2}(c) and \ref{Fig2}(d), the pressure gradients at the receding and the advancing contact line regions are large. By zooming into these regions of the droplet (see figure \ref{profiles_t}) at early times, we observe large interface curvatures, consistent with what one would expect from the results of a large pressure gradient. We want to highlight that as the droplet starts to move across the cone, a film is formed at the receding region. Since the later self-propelling state is independent of the initial conditions, the droplet properties such as the deposited film thickness and the droplet velocity are a function of the droplet position on the cone.

\subsection{Comparison of the numerical solutions of the LAC with molecular dynamics simulations}\label{MD_lub}
 \begin{figure}
\begin{center}
\includegraphics[width=0.9\textwidth]{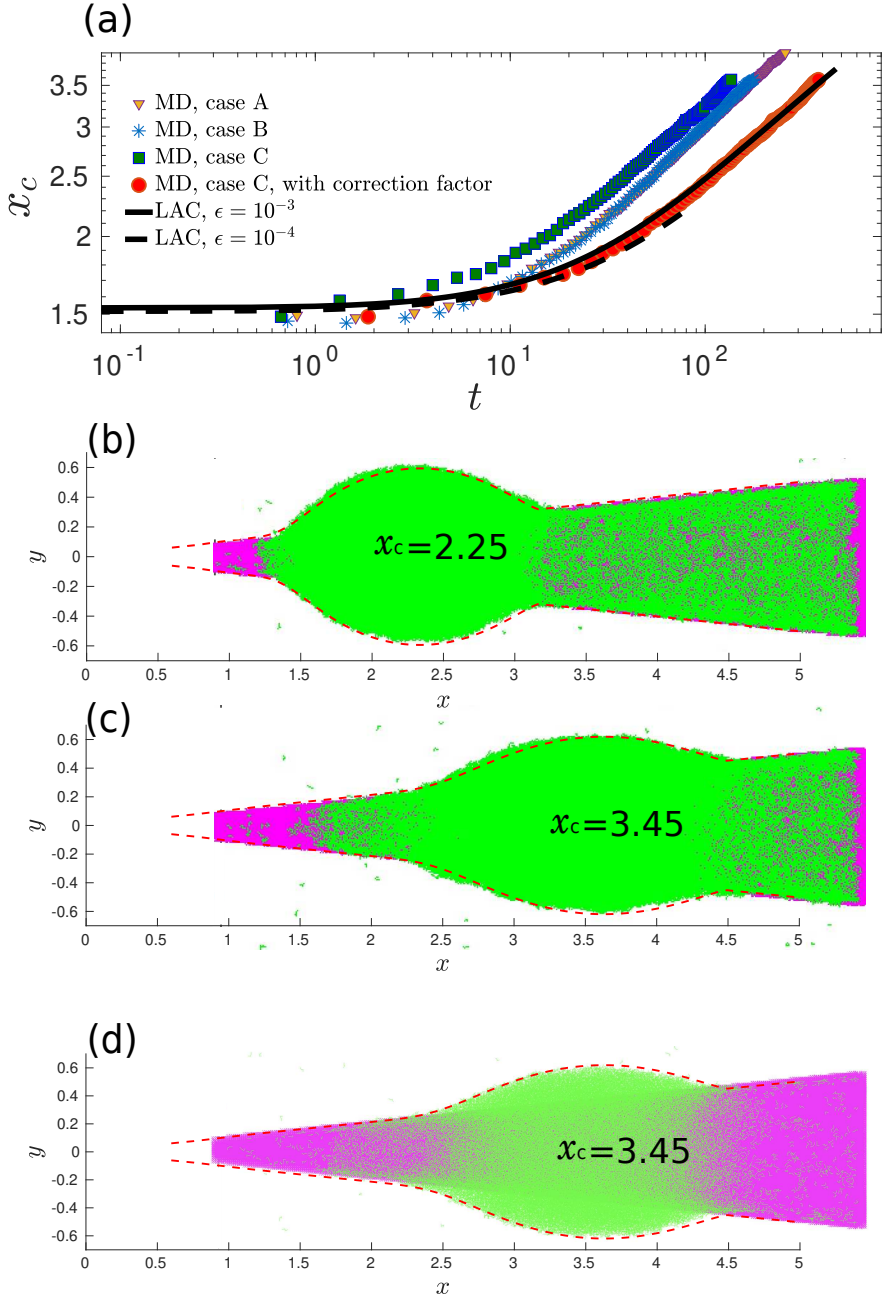}
\caption{(a) The center of mass of the droplet $x_c$ as a function of time $t$ obtained from the MD simulations (symbols) for three different cases and the numerical solutions of the LAC (lines) for  $\lambda=0$ and $\alpha=0.1$ rad. The red circles are results obtained by multiplying the time in case C with a prefactor 2.8. (b), (c) and (d) Comparison between droplet profiles obtained from the LAC and the MD simulations. For each comparison, the profiles are chosen such that   $x_c$ is the same. Both (b) and (c) are for the case C of the MD simulations (wetted substrate) but at two different droplet positions. (d) is for case A (dry substrate) of the MD simulations.  Red dashed curves, profiles from LAC. Green dots: liquid molecules of the droplet. Pink color: the cone substrate.}\label{Fig9}
\end{center}
\end{figure}

We compare the results for the case of cone angle $\alpha=0.1$ rad from the numerical solutions of the LAC and the MD simulations. The comparison serves also as a verification of our lubrication model and can help reveal nanoscopic physical effects. We first compare the center of mass of the droplet $x_c$ as a function of time in figure \ref{Fig9}(a). For the LAC, we have used two different values of prewetted layer thickness, \textit{i.e.} $\epsilon=10^{-4}$ and $10^{-3}$, to highlight their weak influence on the results. We have three cases for the MD simulations. We see that droplets on the dry surface (case A and case B) move slower than the droplet on the wet surface (case C, with the rescaled thickness of the prewetted layer = 0.065). This is consistent with the expectation that the wetted layer reduces the frictional force between the droplet and the substrate. When comparing with the LAC results, we find that the results from the MD simulation (for both the dry surface and wet surface cases)  have a larger non-dimensional velocity. When we multiply the time scale in MD by a prefactor of 2.8 for case C (with a wetted layer), we effectively shift the data from MD horizontally to the right and obtain the results represented by red circles, which makes the two models give the same results.  Some possible reasons for the difference in time scales of LAC and MD are, for example, the differences in the hydrodynamic viscosity, finite size effects and the slip length. In fact, the presence of slip in the simulations is rather unclear because the translation velocity of the drop is much smaller than the thermal velocity of the atoms, by a factor of
10$^{-3}$ or less, and it is not possible to resolve the flow field inside
the drop.  However, simulations of the same liquid in shear flow along a planar solid of the same structure as the cone, under otherwise identical conditions, have a velocity field which extrapolates to zero roughly halfway between the innermost liquid and outermost solid atoms.  If one (naturally) identifies the latter point as the liquid/solid boundary then the slip length is at most a small fraction of an atomic diameter, which is essentially zero. The shapes of the droplet are shown in figure \ref{Fig9}(b)-(d), where the same droplet shapes are predicted by the LAC and the MD when comparing for the same center of mass of the droplet. However, no deposited film is observed for all cases in the MD simulations. 

The effects of the thin prewetted layer are illustrated when comparing a drop advancing on a cone at the same center of mass position
($x_c=3.45$) for the wetted case in figure \ref{Fig9}(c) with the dry case in figure \ref{Fig9}(d). The lighter coloring of the liquid region as compared to figures \ref{Fig9}(b), (c) reflects the fact that there are fewer liquid molecules present in the dry case. For the dry case, the advancing meniscus of the drop is irregular
at molecular scales, corresponding to individual molecules hopping to
attractive sites on the surface, which is also the case for wetting drops 
advancing on a dry flat surface \citep{DDKR96}. The receding meniscus region is 
an uneven film as well, zero to two molecules in thickness, and this
behavior is also present in the prewetted case.  The absence of a continuous
trailing film for these drops is surprising because one would expect a
completely wetting liquid to remain in contact with a solid unless removed
by an external force, and the lubrication calculations in this paper incorporate
this assumption. One possible explanation is the finite size of the
simulated droplets, which may not have enough molecules to exhibit all
features of continuum behavior.  A second, more specific explanation
involves the curvature of the surface.  Liquid adjacent to a flat surface
tends to form pronounced layers and, at least for a crystalline solid, there
is an ordered structure in each layer because the molecules favor positions 
in register with the lattice.  High curvature disrupts the usual lattice 
structure and could thereby weaken the liquid-solid attraction.  In this
vein, it is known that solid curvature has a significant effect on slip
lengths, which are controlled by the same interaction \citep{CZK14,GCR16}.

\subsection{Droplet velocity} \label{drop_vel}

%\begin{figure}
%\begin{center}
%\includegraphics[width=1.0\textwidth]{Fig4.eps}
%\caption{(a) The capillary number $Ca$ rescaled by $1/\ln[c/(\epsilon+\lambda)]$ as a function of the center of mass of the droplet $x_c$. Inset: $Ca$ plotted against the $x_c$. The cone angle $\alpha=0.01$ and $c=0.0815$ is a fitting parameter. The value of $c$ depends on the specific geometry of the problem \citep{Snoeijer13}. }\label{Fig4}
%\end{center}
%\end{figure}

\begin{figure}
\begin{center}
\includegraphics[width=1.3\textwidth]{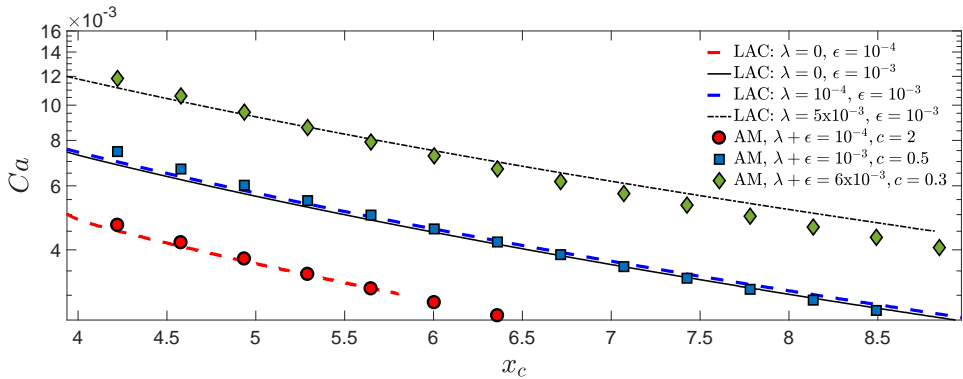}
\caption{Lines: the capillary number $Ca$ as a function of the center of mass of the droplet $x_c$ obtained from LAC. Symbols:  the relation given by (\ref{ca_tha}) with different value of $c$. The cone angle is $\alpha=0.01$. }\label{Fig4}
\end{center}
\end{figure}

 We consider here only cases in which the prewetted layer is much thinner than the droplet, \textit{i.e.} $\epsilon\in [10^{-4}, 10^{-3}]$, but we vary the slip length across a wide range $\lambda \in [0, 20]$. When the prewetted layer is thick, e.g. $\epsilon>10^{-2}$, it becomes unstable quickly due to Rayleigh-Plateau instability \citep{EV08}. 
To investigate the droplet dynamics, we define the capillary number as the dimensionless velocity of the droplet, namely
\begin{equation}\label{xca}
Ca\equiv \frac{dx_c}{dt}.
\end{equation}

\subsubsection{Dependence of droplet velocity on the thickness of the prewetted layer and the slip length}

We start by looking at cases when both the prewetted layer and the slip length are small, \textit{i.e.} $\epsilon\ll1$ and $\lambda\ll1$. In models of dynamical wetting \citep{BEIMR09,Snoeijer13}, these microscopic lengths act as a cutoff length scale for moving contact line singularity and the length scales appear in a logarithmic term of the viscous dissipation \citep{Snoeijer13}. In \cite{chan2020a}, by using asymptotic matching, it is shown that the capillary number scales as $Ca= \theta_a^3/9\ln(c/\lambda)$, where $\theta_a$ is the advancing apparent  contact angle of the corresponding static droplet and $c$ is a fitting parameter. We here propose a similar relation but include the prewetted layer thickness $\epsilon$ as 
\begin{equation}\label{ca_tha}
Ca=\frac{\theta_a^3}{9\ln(c/[\lambda+\epsilon])}.
\end{equation}
With an adjustment of the fitting parameter $c$, this relation describes well the results from the LAC  as shown in figure \ref{Fig4}, but gradually becomes invalid when $\lambda$ is no longer small.

\begin{figure}
\begin{center}
\includegraphics[width=1.0\textwidth]{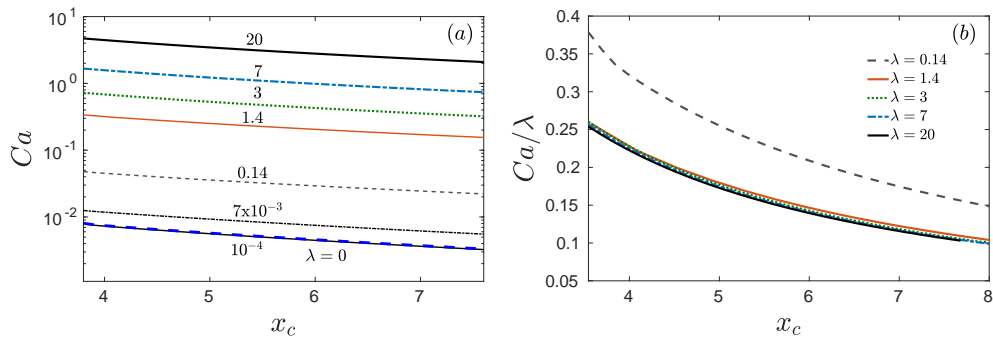}
\caption{(a) The capillary number $Ca$ plotted as a function of the center of mass of the droplet $x_c$ for different slip lengths. (b) The rescaled $Ca$ by $\lambda$ as a function of $x_c$. Parameters: $\epsilon=10^{-3}$ and $\alpha=0.01$ rad. We rescale the data in (a) as $Ca/\lambda$, which collapses the data onto a single curve for $\lambda>1$.}\label{Fig5}
\end{center}
\end{figure}

When exploring a wider range of slip length $\lambda$, one expects a change in the flow profile inside the droplet, from a Poiseuille flow for the case of no-slip to a plug flow as we approach free slip \citep{munch05JEM}. Figure \ref{Fig5}(a) shows $Ca$ as a function of $x_c$ for different slip lengths, where droplets move faster for larger slip lengths as the viscous dissipation is decreased.  At large slip lengths, it is expected that the term with the slip length in the governing equation (\ref{c}) dominates over the other terms, thus $\lambda$ can be scaled out from the equation by defining $\bar{t}\equiv \lambda t$. This implies  $Ca$ scales linearly with $\lambda$. We plot $Ca$ rescaled by $\lambda$ in figure \ref{Fig5}(b), and find that $Ca/\lambda$ collapse onto a single curve for $\lambda>1$, consistent with our expectation. An alternative derivation can also be made based on a balance between the rate of change of capillary energy and the viscous dissipation. The viscous stress scales as $\sim \eta U/\lambda$, giving a bulk dissipation $\sim\eta V U^2/\lambda^2$, which is much smaller than the dissipation due to friction at the substrate $\sim \eta A_w U^2/\lambda$, here $A_w$ is the wetted area. By balancing the dominant viscous dissipation with the rate of change of the surface energy $\partial(\gamma A_w)/\partial t \sim \gamma x_c\alpha U$ gives $U\sim\lambda$.

\subsubsection{Dependence of droplet velocity on the cone angle and the droplet position}

\begin{figure}
\begin{center}
\includegraphics[width=1.15\textwidth]{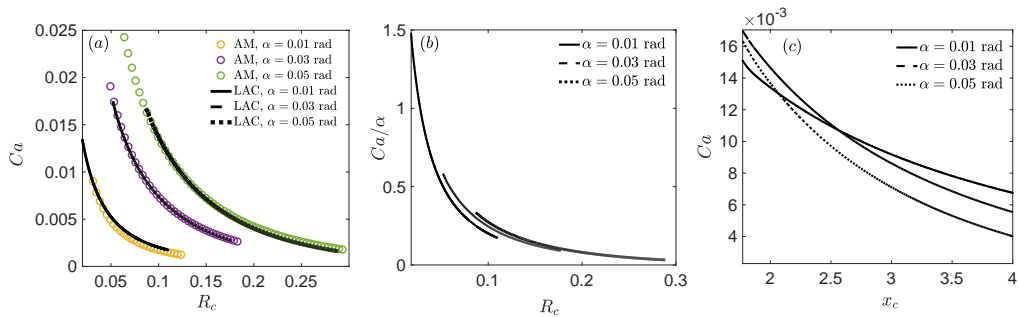}
\caption{Lines: results from LAC. Symbols: the relation given by (\ref{ca_tha}) with $c=2$. The droplet capillary number $Ca$ plotted as a function of $R_c\equiv x_c\tan\alpha$ in (a), and as a function of the center of mass of the droplet $x_c$ in (c). (b) $Ca$ rescaled by $\alpha$ plotted as a function of $R_c$. Parameters: $\epsilon=10^{-3}$ and $\lambda=0$.}\label{Fig6}
\end{center}
\end{figure}

As there is no directional spreading when $\alpha=0$, it is natural to expect that a droplet moves faster at larger cone angles. This is true when comparing $Ca$ at the same cone radius, as shown in figure \ref{Fig6}(a) in which $Ca$ is plotted as a function of $R_c\equiv x_c\tan\alpha$. The results from the LAC agree nicely with the matching results of eq. (\ref{ca_tha}) for the three different values of $\alpha$ using the same value of $c=2$. Remarkably, the agreement is good even when the apparent contact angle is as large as $\theta_a\approx 1$ rad, for example when $R_c\approx 0.06$ and $\alpha=0.03$ rad. Another feature we observe is that $Ca$ decreases when the droplet is at a position of larger cone radius for a fixed cone angle, namely the droplet slows down when moving to the thicker part of the cone. When plotting $Ca$ rescaled by $\alpha$ in figure \ref{Fig6}(b), the results for the three different cone angles nearly collapse onto a single curve.

We have shown that the cone angle and the cone radius give opposite effects to the droplet velocity. It might be interesting to see how the droplet velocity depends on the distance from the tip of the cone, particularly the length of a fibre can be a more important parameter for certain functionality. In figure \ref{Fig6}(c), we show $Ca$ as a function of the droplet's center of mass $x_c$. Remarkably, a non-monotonic behavior is observed. In the limit of large distances from the tip, droplets on  cones with smaller cone angles move faster when comparing at the same $x_c$. When decreasing $x_c$, there are changes of relative strength of $Ca$. For example, at $x_c=2.3$, $Ca$ for $\alpha=0.03$ rad is even higher than that for $\alpha=0.01$ rad. The reason for the non-monotonic behavior is that two factors are playing roles when comparing at the same $x_c$, namely $\alpha$ and $R_c$. The influence of the cone radius $R_c$ is dominant over the cone angle effect when $x_c$ is large, thus droplets move faster at smaller $\alpha$. The cone angle effect becomes more important when $x_c$ is small. Our results demonstrate that a sensitive control of the geometrical factors such as $\alpha$ and $R_c$ is necessary for optimal droplet transport on theses structures.

\subsection{Film deposition} \label{dep_film}
A film is formed at the receding region of the droplet when the droplet moves to the thicker part of the cone, as already shown in figure \ref{profiles_t}(a). We refer to the region that connects the prewetted layer and the deposited film as the film edge region. One can observe from figure \ref{profiles_t}(a) that the film edge region (around $r=1$) propagates much slower than the motion of the droplet. Hence a long deposited film is generated and the film profile is found to remain steady within the simulated time. However, the film would eventually become unstable due to the Rayleigh-Plateau instability, but the time scale for the growth of the disturbance is here greater than the time for the droplet to spread across the cone. For films of nanometric thickness, they can be stabilized by intermolecular forces \citep{QMB1990}.

%When further increasing the cone angle, a different behavior of film deposition is found. A typical example is shown in figure \ref{film_alpha}c for $\alpha=0.1$ rad. The film edge region propagates as fast as the droplet movement, hence a long deposited film is not observed.

\subsubsection{Dependence of the deposited film on the cone angle}
\begin{figure}
\begin{center}
\includegraphics[width=1.7\textwidth]{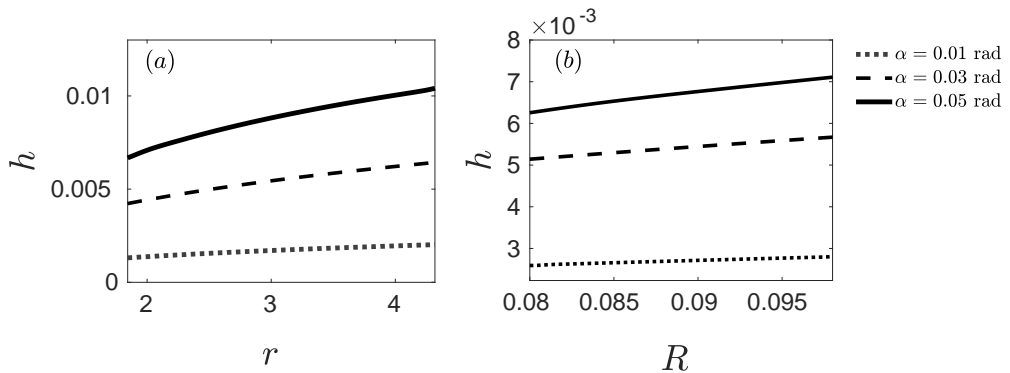}
\caption{(a) The profiles of the deposited film $h$ plotted as a function of the distance from the tip of the cone $r$ for $\alpha=[0.01,0.03,0.05]$ rad. (b) The profiles of the doposited film $h$ plotted as a function of the cone radius $R$. Parameters: $\epsilon=10^{-3}$ and $\lambda=0$.}\label{Fig7a}
\end{center}
\end{figure}

We first consider cases of no-slip ($\lambda=0$). The profiles of the deposited films are shown in figures \ref{Fig7a}(a) for $\alpha=[0.01$, $0.03$, $0.05]$ rad. It is found that the film thickness increases with both $\alpha$ and $r$. It is also important to understand the influence of the cone angle on the film thickness when comparing at the same cone radius. We hence plot in figure \ref{Fig7a}(b) the profiles of the films as a function of $R\equiv r\sin\alpha$. The film is thicker for larger cone angles. As will be explained in section \ref{asymp}, this is mainly due to the larger capillary number for larger cone angles.

\subsubsection{Dependence of the deposited film on the slip length}

\begin{figure}
\begin{center}
\includegraphics[width=1.7\textwidth]{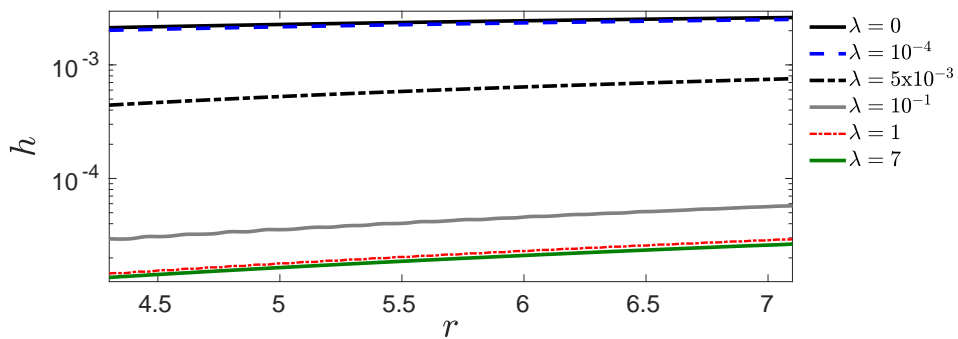}
\caption{The profiles of the deposited film $h$ plotted as a function of the distance from the tip of the cone $r$ for different slip lengths. Parameters: $\epsilon=10^{-3}$ and $\alpha=0.01$ rad.}\label{Fig7b}
\end{center}
\end{figure}

 As droplets move faster at larger slip lengths, one may expect that a thicker film is deposited according to the LLD model. However, our analysis show the opposite results (figure \ref{Fig7b}) with $\alpha=0.01$ rad and $\epsilon=10^{-3}$. The film thickness decreases with the slip length. We find two asymptotic film profiles. One is for the limit of small slip length ($\lambda<10^{-4}$).  Another one is for the limit of large slip length  ($\lambda>1$), which can be understood by the argument that the slip length is absorbed into the time variable $\bar{t}$ as explained in section \ref{drop_vel}. Hence the droplet profile and the deposited film thickness become independent of the slip length. A dramatic change of film thickness appears for slip length in between $10^{-4}$ and $1$. The difference of film thickness between these two limits is of two orders of magnitude. Our results show that when the droplet size is too small, film deposition is not possible as the film thickness computed from our model can be of sub-molecular size. This is particularly relevant for large slip lengths, for which the deposited film is much thinner and the large slip regime can be realized usually for droplet size of micrometers or below.

   %When a droplet is moving faster, it is expected that more fluid at the receding region is left behind, and hence a thicker deposited film is formed. Our results in figure .. show that the dependence on $Ca$ is more complex, we here present a rigorous way to understand the mechanism.

\subsection{Asymptotic matching}\label{asymp}
\begin{figure}
\begin{center}
\includegraphics[width=2.3\textwidth]{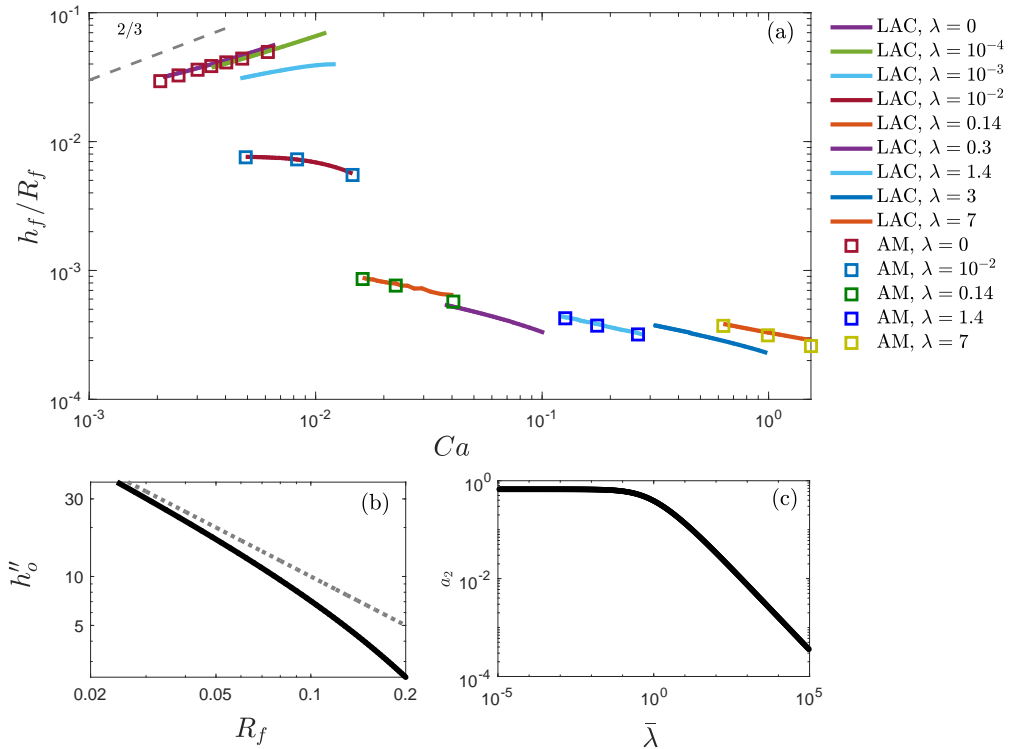}
\caption{(a) The rescaled film thickness $h_f/R_f$ as a function of the capillary number $Ca$ for different slip lengths. Lines: results from the lubrication approach (LAC). Parameters: $\epsilon=10^{-3}$ and $\alpha=0.01$ rad. Squares: results from the asymptotic matching (AM).  The grey dashed line indicates a scaling $Ca^{2/3}$. (b) Solid line: the second derivative at the apparent receding contact line of a static droplet $h''_o\equiv h''_s(r=r_{r})$ as a function of the cone radius of a deposited film $R_f$. The dashed line represents the asymptotic relation $h''_o=1/R_f$ when $R_f\ll 1$. (c) The value of $a_2$ obtained from the 2D lubrication equation for the film region as a function of the rescaled slip length $\bar{\lambda}\equiv \lambda/h_f$. }\label{Fig8}
\end{center}
\end{figure}

 Although we have shown the interface profiles $h(r)$ of the deposited films (figure \ref{Fig7a} and figure \ref{Fig7b}), it is not clear yet how a particular film thickness is related to the capillary number. For droplets spreading on a cone, the motion of the droplet is self-propelled, and the capillary number is a function of the droplet position $x_c$ (or time $t$). Nevertheless, at each moment in time, the droplet deposits a portion of film with a particular thickness $h_f=h_f(x_c)$. Hence we can link a particular film thickness to the corresponding $Ca$ at each droplet position on the cone. The procedure of determining $h_f$ is given in Appendix \ref{appenA}. A natural way to rescale the film thickness $h_f$ is by using the corresponding cone radius $R_f\equiv r_r\sin\alpha$ where the film is deposited. The rescaled $h_f/R_f$ is plotted as a function of $Ca$ for a wide range of slip length $\lambda$ in figure \ref{Fig8}(a) with log-log axes. First, the $2/3$ scaling is not observed for any cases, even for the no-slip case which we would expect from the LLD model. Second, the local slope (in log scales) decreases with the slip length, and becomes negative when the slip length $>10^{-2}$ .

To understand better the numerical solutions of the LAC, we revisit the approach of asymptotic matching. We consider two regions of the liquid-air interface profile: the film region and the static droplet region, which are described by two different force balance equations. We then match the asymptotic profiles of these two regions to determine the deposited film thickness.

\begin{figure}
\begin{center}
\includegraphics[width=1.9\textwidth]{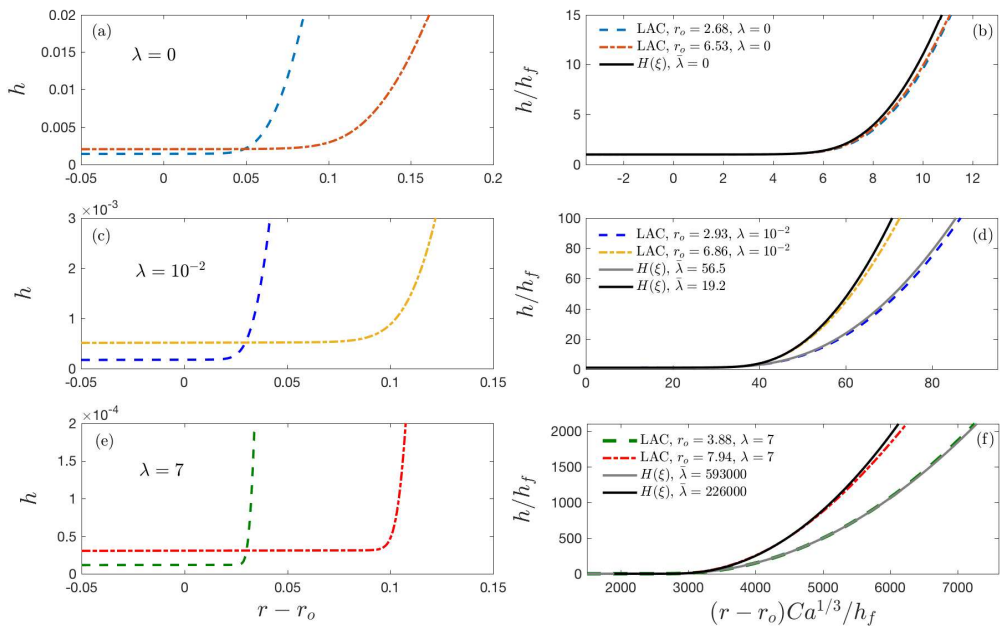}
\caption{(a), (c) and (e): The interfacial profile $h$ as a function of the  shifted radial coordinate $r-r_o$ obtained from LAC in the region connecting the flat film and the droplet for different slip lengths $\lambda$ and droplet positions (characterized by $r_r$ in the right column).  The parameters are: the cone angle $\alpha=0.01$ rad and the prewetted layer thickness $\epsilon=0.001$. (b), (d) and (f): The rescaled profiles $h/h_f$ as a function of $(r-r_o)Ca^{1/3}/h_f$. The solid lines are the numerical solution of (\ref{lld1b4}) with values of $\bar{\lambda}\equiv \lambda/h_f$ computed by the corresponding values of $\lambda$ and $h_f$ from the LAC.}\label{Fig_simSlip}
\end{center}
\end{figure}

As the film thickness is much smaller than the cone radius, we propose that the profile in the film region locally is described by a steady solution $h=h_{2d}(x)$ of the two-dimensional lubrication equation.  In the droplet frame, translating with a velocity $Ca$, the rescaled liquid-air interfacial profile $H(\xi)=h_{2d}/h_f$, here $\xi=xCa^{1/3}/h_f$, follows \citep{SZAFE08}
\begin{eqnarray} \label{lld1b4}
\frac{\partial^3 H}{\partial \xi^3}=\frac{3}{H(H+3\bar{\lambda})}\left(1-\frac{1}{H}\right),
\end{eqnarray}
where $\bar{\lambda}\equiv \lambda/h_f$. We impose a flat film boundary condition $H(\xi\rightarrow -\infty)=1$, and hence close to the flat film, we can write $H=1+\delta \exp[3^{1/3}\xi/(1+3\bar{\lambda})^{1/3}]$, with $\delta\ll1$ \citep{ODB97}. The value of $\delta$ is arbitrary due to the translational invariance of (\ref{lld1b4}).  Here we set $\delta=5 \times 10^{-7}$ when $\xi=0$. When $\xi\rightarrow\infty$, the profile of the film has to match to the droplet shape at the receding region, thus $H$ tends to $\infty$,  the asymptotic solution of (\ref{lld1b4}) is described by $H=a_1\xi+a_2\xi^2$.
The value of $a_1$ and $a_2$ are determined by the numerical solution of equation (\ref{lld1b4}). A comparison between the similarity profile $H(\xi)$ and the rescaled profiles from LAC in the region connecting the flat film and the droplet is given in figure \ref{Fig_simSlip} for three different slip lengths: $\lambda=0$, $10^{-2}$ and $7$.  The profiles from LAC are shifted manually by $r_o$ so that they match the best with the solution of (\ref{lld1b4}). The value of $r_o$ is close to $r_r$ (difference within 2 $\%$ ). Note that for non-zero slip lengths, as $h_f$ varies with droplet positions, $\bar{\lambda}$ has different values at different droplet positions even though $\lambda$ is the same. We can see in figure \ref{Fig_simSlip} that the similarity profiles describe well the profiles from the LAC particularly at the region closer to the flat film. Away from the flat film, the profiles from the LAC bend to match the droplet shape.

In the static droplet region, the profile $h_s(r)$ is determined by the static equation of uniform curvature $\kappa_s$ obtained by substituting $h_s(r)=h(r)$ into Eq.\eqref{cur}, with a magnitude of $\kappa_s$ that depends on the droplet position on the cone.
The problem is closed by including the boundary conditions $h_s(r=r_{r})=0$ and $h_s'(r=r_{r
})=0$ at the substrate.
%\begin{equation}\label{dd}
%\kappa_s\equiv\frac{h_s^{''}}{(1+h_s^{'2})^{3/2}}-\frac{1}{(r\alpha
%	+h_s )\left[1+\left(\frac{h_s'+\alpha}{1-h_s'\alpha }\right)^{2}\right]^{1/2}}=K,
%\end{equation}

Now we are in a position of matching the two asymptotic profiles in the overlapping region. As we already impose the condition $h_s'(r=r_{r})=0$ for the droplet region. A natural matching condition is equating the second derivatives of the asymptotic profiles. In the film region,
\begin{equation}\label{filmr1}
h''_{2d}(\xi\rightarrow \infty)=2a_2 Ca^{2/3}/h_f.
\end{equation}
Matching  $h''_{2d}(\xi\rightarrow \infty)$ to the second derivative $h''_o\equiv h''_s(r=r_{r})$ in the static droplet region gives the film thickness
\begin{equation}\label{hf}
h_f=\frac{2a_2(\bar{\lambda})}{h''_o(R_f)}Ca^{2/3}.
\end{equation}
Importantly, as all the cone angles are small, $h''_o$ is independent of $\alpha$ but only a function of the cone radius $R_f$, which is plotted in figure \ref{Fig8}(b) in log-scales. Note also that $a_2$ is determined from (\ref{lld1b4}) and a function of $\bar{\lambda}\equiv \lambda/h_f$, which is plotted in figure \ref{Fig8}(c). We note that $h''_o=1/R_f+\kappa_s$, and thus $h''_o\approx 1/R_f$ when $R_f\ll 1$, which is represented by the dashed line in figure \ref{Fig8}(b). With the computed values of $h''_o$ and $a_2$, and using the values of $Ca$ for each droplet position obtained from LAC, we plot $h_f/R_f$ computed from (\ref{hf}) in figure \ref{Fig8}(a) as square markers. Remarkably, the results from the asymptotic matching agree with the numerical results for all slip lengths. Provided the excellent agreement between the two approaches, we can understand our results in terms of the flow inside the film and the geometry of the droplet, and hence provide a better picture of the physical mechanism of film deposition by a droplet moving on a cone.

\begin{figure}
\begin{center}
\includegraphics[width=1.9\textwidth]{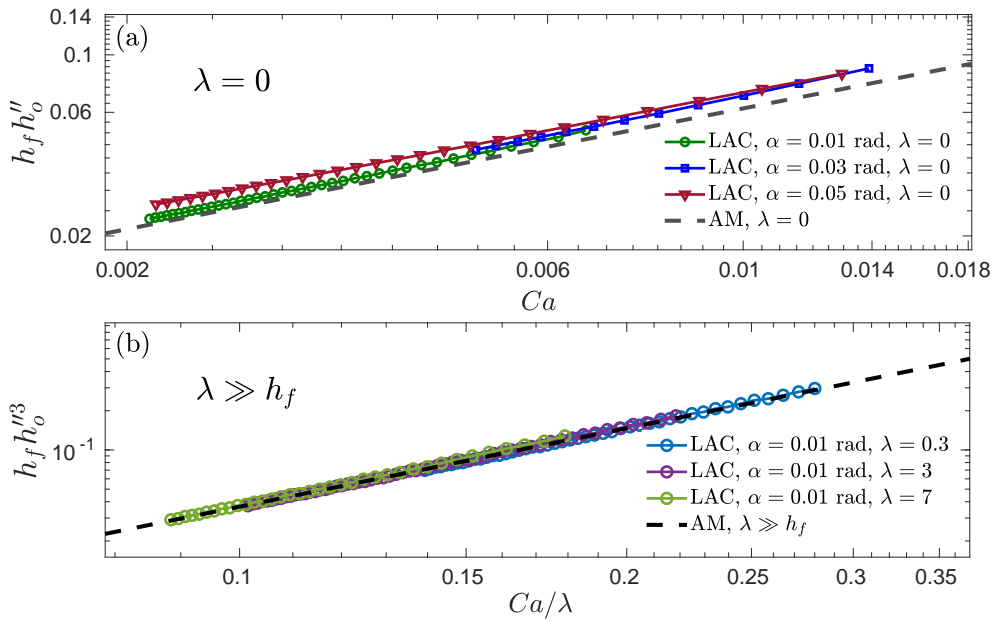}
\caption{(a) The film thickness $h_f$ rescaled by $1/h''_o$ as a function of $Ca$ for $\lambda=0$. Symbols are results from LAC for three different cone angles. The prewetted layer thickness $\epsilon=10^{-3}$. The dashed line is the result from AM: $h_fh''_o=1.34 Ca^{2/3}$. (b) The film thickness $h_f$ rescaled by $1/h''^{3}_o$ as a function of $Ca/\lambda$ for three different slip lengths $\lambda\gg h_f$. Symbols are results from LAC. The dashed line is the result from AM: $h_fh''^{3}_o=3.667 (Ca/\lambda)^{2}$.}\label{Fig8_2}
\end{center}
\end{figure}

We first look at the no-slip case. When $\bar{\lambda}=0$, $a_2=0.669$, which is the same value as  obtained from previous studies \citep{RIO2017100}. The description of the film region is the same as, for example, the dip-coating cases. Then why is the $2/3$ scaling not obtained when plotting $h_f/R_f$ as a function of $Ca$? One important aspect in our problem is that the droplets have a finite size. Hence there are two length scales: the cone radius and the droplet radius. In terms of rescaled quantities, this means that the second derivative $h''_o$ is not a linear function of $1/R_f$, except when $R_f\ll1$, which is already demonstrated in figure \ref{Fig8}(b). When we rescale the results of $h_f$ obtained from the LAC by $1/h''_o$ and plot it as a function of $Ca$ in figure \ref{Fig8_2}(a) for three different cone angles. The scaling $Ca^{2/3}$ is recovered and agrees well with the prediction from asymptotic matching especially for smaller cone angles.

Next we look at the slip dependence. From figure \ref{Fig8}(c), we see $a_2$ is independent of $\bar{\lambda}$ when $\bar{\lambda}\ll1$.  Hence for $\lambda \ll h_f$, the film thickness becomes independent of $\lambda$. As the typical order of magnitude of $h_f$ for a no-slip case is $10^{-3}$, $h_f$ starts to depend on $\lambda$ significantly when $\lambda>10^{-3}$. This is consistent with our numerical results. For $\bar{\lambda}\gg1$, we find that $a_2=0.771\bar{\lambda}^{-2/3}$. Substituting this expression of $a_2$ into (\ref{hf}), we obtain
\begin{equation}\label{hf_slip}
h_f=\frac{3.667}{h''^{3}_o}\left(\frac{Ca}{\lambda}\right)^2.
\end{equation}
This expression is in perfect agreement with our numerical results from the LAC for $\lambda\gg h_f$ in figure \ref{Fig8_2}(b). For droplets moving on a conical fibre, we show already that $Ca \sim \lambda$ when $\lambda\gg1$, equation (\ref{hf_slip}) then suggests $h_f$ is independent of the slip length as shown in figure \ref{Fig7b} for $\lambda\gg1$. 

 One may expect that the droplet profile does not maintain quasi-static shape when  the slip length is not small due to significant viscous effects over the entire droplet. However, the excellent agreement between the results from the lubrication equation on a cone and the approach of asymptotic matching suggests that the quasi-static assumption is still valid. The reason might be the large length separation between the deposited film and the droplet height maintaining a very large difference in time scales, as one can observe from the mobility term which scales as $\sim\lambda h^2$. Thus there is sufficient time for the droplet to relax to a quasi-static shape when the apparent contact lines move. For the large slip length regime, elongational flow has been proposed to appear and dominate the viscous dissipation  \citep{munch05JEM}, which has been observed for dewetting droplets \citep{McGraw16,chan2017}. The large slip regime in our model, assuming Poiseuille flow as the dominating flow structure, is considered as the intermediate slip regime in the analysis of \cite{munch05JEM}. For a translating droplet, the effect of elongational flow is unclear, which requires additional experimental and theoretical studies.

 For partially wetting surfaces, the droplet dynamics is the same as for the perfectly wetting cases if the prewetted layer is thick enough so that van der Waals forces between the liquid-air and the solid-liquid interfaces can be neglected. A partially wetting droplet moving on a cone without a prewetted layer will also deposit a film if it moves with a velocity above a critical value. The properties of the film are expected to be similar to the wetting cases, namely following the asymptotic relation (\ref{hf}), as long as the droplet maintains an axisymmetric shape.

\section{Conclusions}\label{dis_con}
The directional spreading of a viscous droplet on a conical fibre due to capillarity is investigated for small cone angles and for a wide range of slip lengths by using the lubrication equation on a cone. The droplet velocity increases with the cone angle and the slip length, but decreases as the cone radius becomes larger. At the receding part of the droplet, a film is deposited on the cone surface while the droplet is moving. When comparing with our MD simulations, we find that the droplet shapes obtained from these two approaches are the same. The velocity also shows a similar trend. However, no deposited film is observed in the MD simulations, which might be due to the nanoscopic  size of the droplet.    

The thickness of the deposited film observed in the LAC decreases from $h_f\approx10^{-3}$ for the no-slip case ($\lambda=0$) to $h_f\approx 10^{-5}$ for $\lambda> 1$.
We show that the film thickness obtained from the lubrication model can be understood by a similar approach of asymptotic matching used in the LLD model. For the no-slip limit, the standard $Ca^{2/3}$ scaling is recovered only when the length scale is given by $1/h''_o$ in the re-scaling. In the limit of $\lambda\gg h_f$, we find another asymptotic regime in which the film thickness scales as $h_f h''^{3}_o\sim (Ca/\lambda)^2$. For the problem we study here, the crossover of these two regimes occurs at $\lambda\approx 10^{-4}-10^{-1}$. Our results show that manipulating the droplet size, the cone angle and the slip length provides  different schemes for guiding droplet motion and coating the substrate with a film.

\section{Acknowledgments}
T.S.C. and A.C. gratefully acknowledge financial support from the UiO: Life Science initiative at the University of Oslo. A.C. is grateful for the financial support from the Norwegian Research Council, project number 263056 and 301138. JK was supported in part by the US National Science Foundation under grant No. 1743794. We thank Dr. Kari Dalnoki-Veress and Carmen Lee for stimulating discussions.

\appendix
\section{Determination of $r_r$, $r_a$, $h_f$ and $R_f$} \label{appenA}
\begin{figure}
\begin{center}
\includegraphics[width=1.2\textwidth]{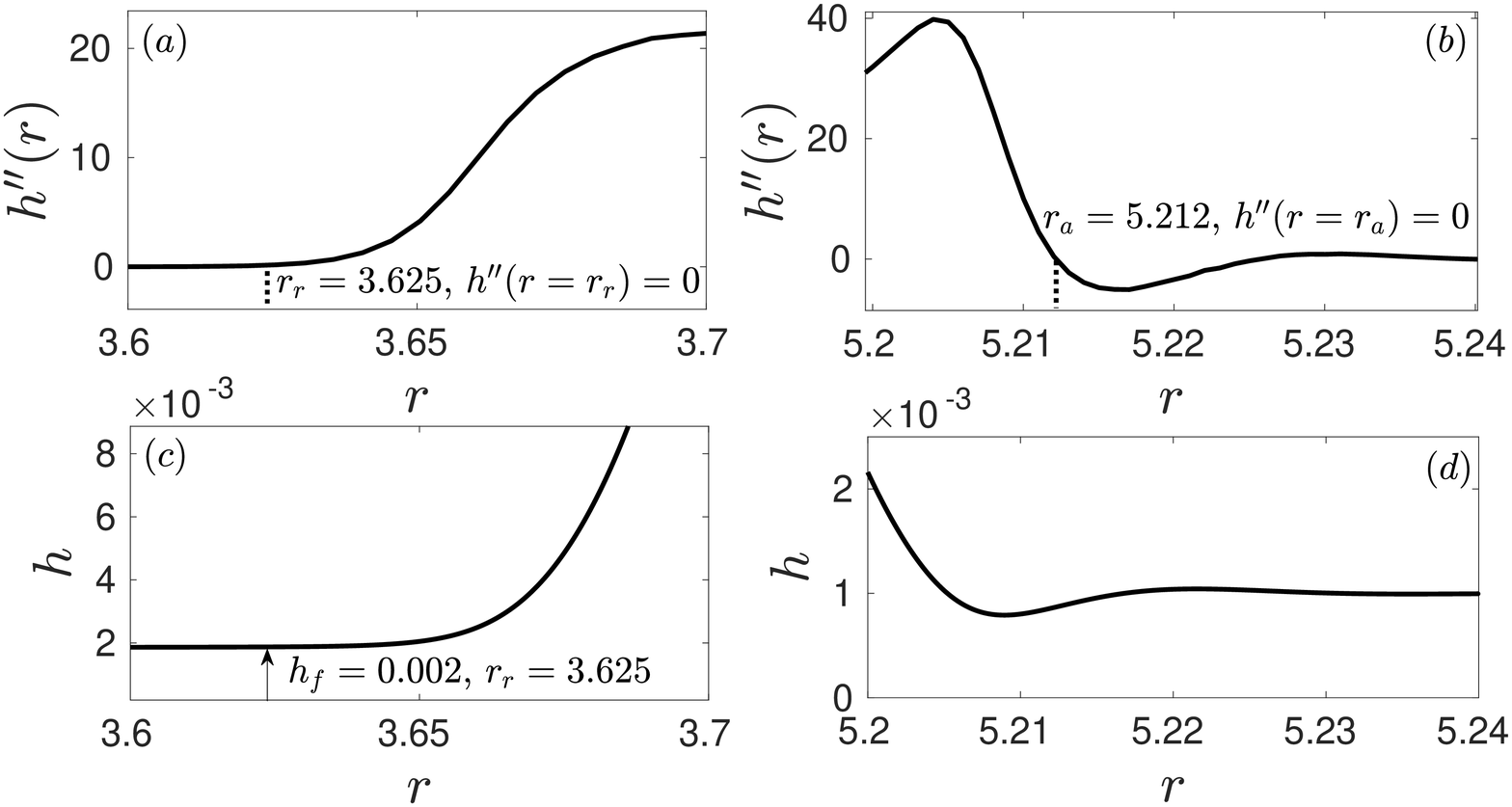}
\caption{(a) The second derivative $h''(r)$ of the liquid-air interfacial profile at the receding region of the droplet that connects to the LLD film at $t=309$. (b) The second derivative $h''(r)$ of the liquid-air interfacial profile at the advancing region of the droplet that connects to the prewetted liquid layer at $t=309$. (c) The liquid-air interfacial profile $h(r)$ at the same range of $r$ as (a). (d) The liquid-air interfacial profile $h(r)$ at the same range of $r$ as (b). Parameters: the cone angle $\alpha=0.01$ rad, the prewetted layer thickness $\epsilon=10^{-3}$ and the slip length $\lambda=0$.}\label{find_hf}
\end{center}
\end{figure}

Since the liquid-air interface of the droplet is continually connected to the liquid-air interface of the prewetted layer/ LLD film, we define the domain of the droplet as follows. We denote the boundaries of the droplet in the receding and the advancing regions respectively as $r=r_r$ and $r=r_a$. We first compute the second derivative of the profile $h''(r)$ at a certain time, which is shown in figure \ref{find_hf}(a) and \ref{find_hf}(b) respectively in the receding and the advancing regions. The second derivative drops to zero when approaching the film regions. Since $h''$ is non-negative in the receding region, the droplet boundary $r=r_r$ is defined as the position at which $h''(r)$ drops to below $0.01$, \textit{i.e.} $h''(r=r_r)=0.01$. In the advancing region, $r=r_a$ is defined as the position at which $h''(r)$ vanishes, \textit{i.e.}  $h''(r=r_a)=0$. The thickness of the deposited film at that particular time is defined as $h_f(t)=h(r=r_r,t)$ and the corresponding cone radius $R_f(t)=r_r\alpha$. Hence we can link the film thickness $h_f$ to the capillary number $Ca$ at each time.
%\frac{\partial h}{\partial t}+\frac{1}{3\eta }\frac{\partial}{\partial r} \left[  h^2(h+3\lambda)\frac{\partial p}{\partial r}\right]+\frac{h^2(h+3\lambda)}{3\eta r}   \frac{\partial p}{\partial r}=0

\section{Dependence on the initial profile} \label{appenB}
\begin{figure}
\begin{center}
\includegraphics[width=1.2\textwidth]{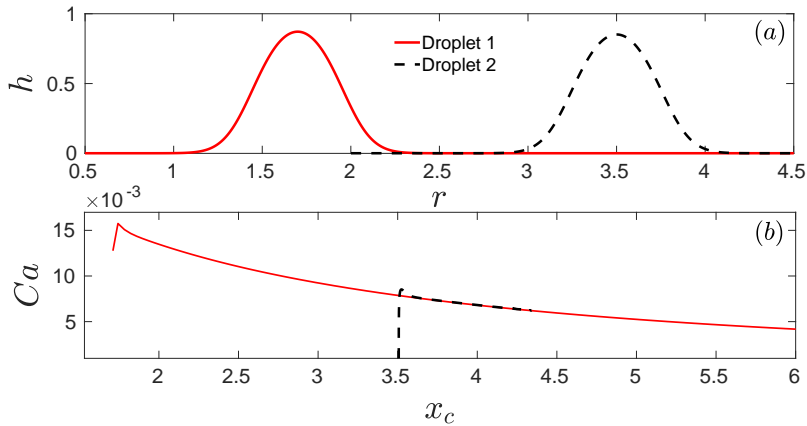}
\caption{(a) Two different initial droplet profiles. The droplets profiles are described in section \ref{num_met} with $\epsilon=10^{-3}$. Droplet 1: $A=0.87$ and $r_i=1.7$. Droplet 2: $A=0.85$ and $r_i=3.5$.  (b) The capillary number $Ca$ as a function of the droplet position $x_c$. Red solid line: droplet 1. Black dashed line: droplet 2. Parameters: $\alpha=0.01$ rad and $\lambda=0$.}\label{dep_init}
\end{center}
\end{figure}

To investigate how the droplet dynamics depends on the initial profile of the droplet, we here compare the dynamics of two droplets with different initial profiles. The initial profiles of the two droplets are described in section \ref{num_met} with different values of $A$ and $r_i$, see figure \ref{dep_init}(a). We compare the capillary number as a function of the droplet position in figure \ref{dep_init}(b). We can see after an initial quick relaxation, the later dynamics is independent of the initial profiles.

\bibliographystyle{jfm}
\bibliography{new_all_ref}

\end{document}